\documentclass[aps,prd,groupedaddress,nofootinbib,letterpaper]{revtex4}

\usepackage{amsmath}
\usepackage{amsfonts}
\usepackage{amssymb}
\usepackage{amsthm}

\usepackage{braket,mathtools}
\usepackage{mathrsfs}
\usepackage{stmaryrd}

\usepackage{hyperref}
\usepackage{color}

\DeclarePairedDelimiter\abs{\lvert}{\rvert}

\newcommand{\call}{{\cal L}}
\newcommand{\calo}{{\cal O}}
\newcommand{\calh}{{\cal H}}

\newcommand{\beq}{\begin{equation}}
\newcommand{\eeq}{\end{equation}}
\newcommand{\bea}{\begin{eqnarray}}
\newcommand{\eea}{\end{eqnarray}}
\newcommand{\hf}{\frac{1}{2}}

\begin{document}
\begin{titlepage}

\title{Gravitational dressing, soft charges,\\  and perturbative gravitational splitting}

\author{Steven B. Giddings}
\email{giddings@ucsb.edu}
\affiliation{Department of Physics, University of California, Santa Barbara, CA 93106}
\affiliation{CERN, Theory Department,
1 Esplande des Particules, Geneva 23, CH-1211, Switzerland}

\begin{abstract}

In gauge theories and gravity, field variables  are generally not gauge-invariant observables, but such observables may be constructed by ``dressing" these or more general operators.  Dressed operators create particles, together with their gauge or gravitational fields which typically extend to infinity.  This raises an important question of how well quantum information can be localized; one version of this is the question of whether soft charges fully characterize a given localized charge or matter distribution.  This paper finds expressions for the non-trivial soft charges of such dressed operators.  However, a large amount of flexibility in the dressing indicates that the soft charges, and other asymptotic observables, are not inherently correlated with details of the charge or matter distribution. Instead, these asymptotic observables can be changed by adding a general radiative (source-free) field configuration to the original one.  A dressing can be chosen, perturbatively, so that the asymptotic observables are independent of details of the distribution, besides its total electric or Poincar\'e charges.  This provides an approach to describing localization of information in gauge theories or gravity, and thus subsystems, that avoids problems associated with nonlocality of operator subalgebras.  Specifically, this construction suggests the notions of electromagnetic or gravitational splittings, which involve networks of Hilbert space embeddings in which the charges play an important role.

\end{abstract}

\maketitle

\end{titlepage}

\section{Introduction}

Quantum information has become an important theme in current theoretical physics.  However, in the context of gauge theories and in particular gravity, there are various puzzles about how to describe it, and in particular its localization.  These puzzles in fact drive at the heart of some of the most challenging questions in physics:  can information be localized in a black hole, and how does it escape?  Or, is that information manifest, in some way, in the gravitational field surrounding the black hole?  Is there a precisely equivalent description of physics in a region of spacetime in terms of variables outside the region, or at an asymptotic boundary, as in anti de Sitter space?  In typical quantum systems, a notion of localization of information, {\it e.g.} in quantum subsystems, is a basic concept that is prior to many others, such as entanglement, information transfer, and computational complexity; is that true in gravity?

A way to investigate these questions is through the study of observables, which can create excitations of the Hilbert space.  Indeed, a careful way to look at subsystems in field theory is to define them via commuting subalgebras of the algebra of observables, associated to spacelike separated regions.\footnote{For treatment of field theory from this perspective, see \cite{Haag}; for this view of subsystems, see \cite{SGalg}.  In this paper, ``observables" mean quantum observables, which ordinarily are described as gauge-invariant self-adjoint operators on the Hilbert space.  These are semantically distinguished from ``things that observers/experimentalists observe", though of course are related.  One can use the terminology ``q-observable" for the former, in cases where confusion might arise; for further discussion see {\it e.g.} \cite{GiSl2}.}  
But, while the field variables are basic observables in non-gauge theories, the field variables are not typically gauge invariant in gauge theory or gravity, hence are not physical observables.  One way to rectify this situation is to ``dress" fields, or more general operators, to construct gauge-invariant observables.  

Such dressed observables, typically create a nontrivial gauge or gravitational field extending to infinity, indicating a basic kind of nonlocal behavior.  This raises the question of how much information about a given charge or mass distribution in a region is accessible in the corresponding field that extends outside the region.
For example, soft charges \footnote{See \cite{astrorev} for a review.} have been proposed as an important characteristic of asymptotic gauge and gravitational fields, and it has been suggested that asymptotically-measured soft charges carry a lot of information about the corresponding charge or matter distribution\cite{StZh,Hawk,HPS1,HPS2,astrorev,astrorevisit,HPPS}.   

This paper will examine these questions.  Indeed, since dressed observables typically (and in gravity, must\cite{DoGi2}) create non-trivial asymptotic fields, we expect them to carry nontrivial soft charges.  This paper studies the form of the soft charges for these operators, for electromagnetism (EM) and for gravity.  It then turns to the question of how much these soft charges depend on the charge or matter configuration.  In fact, one can see that there is a lot of flexibility in choosing the dressing of a given operator or set of operators, and correspondingly a given operator can have many different asymptotic fields and soft charges.\footnote{This includes asymptotic configurations not satisfying antipodal matching.  Dressings are also important in proper treatment of IR effects in scattering, as seen for example in \cite{Kapec:2017tkm,Choi:2017ylo}.}  This dressing ambiguity corresponds to the flexibility to add an arbitrary radiative (source-free) field on top of a given field configuration.  One can in fact show, in EM and gravity, that the only required correlation of the field outside a region and the distribution inside the region is through the total electric charge (magnetic charge is neglected), or the total Poincar\'e charges, respectively.  In this sense, the soft charges, and indeed other EM or gravitational observables, can be ``decoupled" from the charge or matter distribution.\footnote{For related observations see \cite{BoPo}.}  Put differently, the soft charges characterize features of the radiation field that has been superposed on any original field, rather than characterizing the charge or matter distribution.  This plausibly also holds for matter in a black hole, suggesting that soft charges will not help with the problem of unitarity.

This construction in fact provides a way of describing localization of information in gauge theory, and in gravity, at least a perturbative level.  While subsystems are not easily defined any longer in terms of commuting subalgebras, it appears that they can be defined in terms of certain embeddings of product Hilbert spaces into the full Hilbert space, which we refer to as electromagnetic or gravitational splittings\cite{DoGi3,QFG,DoGi4}.  While this construction is particularly clear in the EM case, some puzzles remain in the gravitational case, and in particular the question of nonperturbative completion of this structure is an important one\cite{QFG}.

\section{The electromagnetic case}

An important question in EM is how much we can learn about a  charge distribution in a region by making observations of the EM field of that distribution outside that region.  Specifically, in making contact with quantum information questions, we would like to understand to what extent information is localized -- can we define a localized ``electromagnetic qubit"? The reason information isn't completely localized is because of the EM constraints, which enforce gauge invariance and imply a certain amount of nonlocality.

\subsection{Dressed operators}

We investigate this question by considering a scalar $\phi(x)$, coupled to EM with charge $q$,
\beq
\call = -\frac14 F^{\mu \nu} F_{\mu \nu} -\abs{D_\mu \phi}^2 - m^2 |\phi|^2\ .
\eeq
Here $F_{\mu\nu}=\partial_\mu A_\nu -\partial_\nu A_\mu $, $D_\mu \phi = \partial_\mu\phi - iqA_\mu \phi$, and gauge transformations act as 
\beq\label{gaugexm}
\phi(x) \to e^{-i q \Omega(x)} \phi(x)\quad,\quad A_\mu \to A_\mu -\partial_\mu \Omega .
\eeq

The local field $\phi(x)$ is not gauge invariant, and therefore is also not an observable.  A corresponding gauge-invariant can be constructed, however, by dressing $\phi(x)$.  A particularly simple dressing is the Faraday line\cite{Dirac1955},
\beq\label{Fline}
\Phi_\Gamma(0,\vec x)= \phi(0,\vec x)e^{iq\int_\Gamma A}\ ,
\eeq
where $\Gamma$ is an arbitrary curve connecting $\vec x$ to infinity within the $t=0$ slice.  This is clearly invariant under \eqref{gaugexm}, for a gauge transformation with $\Omega$ vanishing at infininty.  Equivalently, one can show that this operator commutes with the EM constraint,
\beq
\partial^\mu F_{0\mu} - j_0=0
\eeq
where $j_\mu$ is the EM current; the constraint  generates gauge transformations.
Concrete expressions for commutators can be worked out as in \cite{DoGi1} by, {\it e.g.}, employing a Feynman gauge-fixing term,
\beq
\call_{\rm gf}=- \frac{1}{2\alpha} (\partial_\mu A^\mu)^2\ .
\eeq

Expression \eqref{Fline} of course does not describe the most general gauge-invariant dressing.  A much broader class of dressings is of the form
\beq\label{gdress}
\Phi_{\check E}(0,\vec x)= \phi(0,\vec x)e^{i\int d^3x' {\check E}_{\vec x}^i(\vec x') A_i(0,\vec x')}\ ,
\eeq
where ${\check E}^i$ is a classical electric field satisfying the equation
\beq
\nabla'_i{\check E}^i_{\vec x}(\vec x') = q \delta^3(\vec x'-\vec x)\ .
\eeq
These are easily seen to also be gauge-invariant, and when applied to the vacuum, \eqref{gdress} creates a $\phi$ (anti)particle together with an electric field $-{\check E}^i_{\vec x}$.

Clearly there is an infinite number of such dressings.  The states they create differ from one another in the radiative (source-free) part of the EM field.  For example, the operator \eqref{Fline} creates an electric field concentrated in a thin filament.  This is not the most energetically favored field state, and so will emit radiation to null infinity (${\cal I}^+$) and, for a static source in the absence of other influences, will settle down to
a Coulombic configuration\cite{Shab,PFS,HaJo}.  

It is also worth noting that while special configurations with antipodal identification of the field at infinity have been considered\cite{HMPS}, no such identification is assumed here.  The dressing   \eqref{Fline}, where $\Gamma$ is the positive $z$ axis, illustrates this.  Of course, regulation is needed in order to avoid an infinite energy configuration.  An example of such a regulated dressing arises from
\beq\label{conefield}
{\check E}^r= q\frac{f(\theta^A)}{r^2}\quad ,\quad {\check E}^A=0\ .
\eeq
where $f(\theta^A)$ is a function of the angular coordinates $\theta^A$, satisfying $\int d\Omega f=1$.  For example, $f$ 
can be taken as a window function selecting a small angular region around the positive $z$ axis.  Thought of as a classical configuration, \eqref{conefield}, together with $\vec B=0$, are finite-energy initial data for the EM field, which therefore have consistent evolution.  The finite energy condition at large $r$ follows immediately from the expression for the EM energy.  There is a divergence at $r=0$, but by smoothly transitioning the field lines near $r=0$ to those of the Coulomb configuration, this divergence becomes simply the usual Coulombic one, and the energy difference from that is finite.  Thus, in particular, this construction provides an argument that general EM field configurations need not be antipodally identified at infinity.\footnote{The field \eqref{conefield} can yield a logarithmically-divergent boost charge, or center of energy, $M_{0i}$.  However, it is not clear that this leads to pathology.  Thanks to M. Henneaux for suggesting examination of this point.}

\subsection{Soft charges}

Soft charges provide an interesting set of characteristics of an EM field (for a review, see \cite{astrorev}).  These can be defined either at null infinity, $\cal I^{\pm}$, or at spatial infinity, $i^0$.  For a given function $\epsilon(\theta^A)$, the soft charge at $i^0$ is 
\beq
Q_\epsilon= \int d\Omega\, \epsilon(\theta^A) \lim_{r\rightarrow\infty}\left( r^2 E^r\right)\ ,
\eeq
and similarly at $\cal I^{\pm}$.

Since dressed operators create nontrivial asymptotic fields, they should have nonzero soft charges.  Indeed, we easily find
\beq\label{softcomm}
[Q_\epsilon,\Phi_{\check E}] = -\left[ \int d\Omega\, \epsilon(\theta) \lim_{r\rightarrow\infty} \left(r^2 {\check E}^r\right)\right]\ \Phi_{\check E}\ ;
\eeq
in the special case of \eqref{conefield} (or of asymptotic behavior of this form),
\beq
[Q_\epsilon,\Phi_{\check E}] = -q\left[\int d\Omega\, \epsilon(\theta) f(\theta)\right]\ \Phi_{\check E}\ .
\eeq
In accord with the previous discussion, we easily see that operators can be constructed with arbitrary values of the soft charges, and in particular without necessity of antipodal identification.

One can also describe the soft charges of a state with multiple dressed particles,
\beq
|\Psi\rangle = \Phi_{\check E_1}\cdots\Phi_{\check E_n}|0\rangle\ ;
\eeq
we find
\beq\label{chargevalue}
\langle\Psi| Q_\epsilon |\Psi\rangle = \langle \psi |Q_\epsilon|\psi\rangle + \langle\psi |\psi\rangle \sum_{k=1}^n Q_{\epsilon,k}
\eeq
with 
\beq
|\psi \rangle = \phi_1\cdots\phi_N |0\rangle
\eeq
and with
individual charges $Q_{\epsilon,k}$ as in \eqref{softcomm}.\footnote{Na\"\i vely the first term of \eqref{chargevalue} can be set to zero; however this isn't necessarily true if there is an infinite number of vacua acted on by the soft charges\cite{BBM,Sachs}\cite{astrorev}.  We leave investigation of this for separate work.}
This expression also generalizes if one considers a product of soft charges.

\subsection{Decoupling soft charges}

The pieces are now in place to ask what information can be determined about a charge distribution, say localized in a neighborhood $U$, by measurements of the EM soft charges.  Consider in particular states created by a collection of $N$ dressed operators of the form \eqref{gdress}, where the $N$ $\vec x_k$ lie in $U$.  If we pick a specific ${\check E^i}$ that is used to dress all the particles, then field observations outside $U$ can measure aspects of the charge distribution inside $U$; for an example consider dressing all the particles by Faraday lines, \eqref{Fline}.  However, this particular choice of dressing of the operators assumes a particular correlation between the state of the EM field and the configuration of particles.  There are equally good dressings that do not have this correlation.  

For example, another dressing of the combined operator $\phi(\vec x_1)\cdots \phi(\vec x_N)$ can be found by choosing a fiducial point $\vec y\in U$, and defining the dressed operator
\beq\label{commdress}
e^{iN\int d^3x' {\check E}_{\vec y}^i(\vec x') A_i(0,\vec x')} \prod_{k=1}^N\left[ \phi(0,\vec x_k)e^{iq\int_{\Gamma_k} A}\right]\ ,
\eeq
where $\Gamma_k$ are curves from $\vec x_k$ to $\vec y$, within $U$, and ${\check E}_{\vec y}^i$ is some arbitrarily-chosen ``standard dressing" with  charge $q$.  
This operator creates an EM field outside $U$ that depends only on the choice of the form of the standard dressing, and on the total EM charge $Q=Nq$; this also means that the soft charges are determined by this choice, and by $Q$. 
Once again, the dressing field created by \eqref{commdress} differs from that of the preceding dressing, or from {\it e.g.} a collection of Coulomb fields, by a purely radiative (sourceless) configuration of the EM field.

These statements lead to the following conclusions:

\begin{enumerate}

\item  Measurements at $i^0$ (and those of soft charges, in particular) can't necessarily detect details of the charge configuration (state) inside $U$.  

\item  Such measurements {\it do} detect properties of the radiation field that has been superposed on this charge configuration.

\item  We can always {\it choose} to consider initial states where the charge distribution and the radiation field are correlated in a certain way, but we can also choose states where they are correlated in a different way.  The only necessary correlation is determined by the total charge $Q$; by Gauss' law, the asymptotic EM field must have a total EM charge $Q$.

\item  In states with specific correlations between the charge distribution and the asymptotic field, one can of course asymptotically measure those aspects of the charge distribution correlated with the asymptotic field.

\end{enumerate}

In this sense, the soft charges measured at infinity are in general independent of any details of the charge distribution.  This extends to more general asymptotic measurements of the EM field.  Practically, it means that information can be localized ({\it i.e.} is not measurable outside $U$) for  charged states in QED, or put colloquially, there can be ``electromagnetic qubits."

\subsection{Electromagnetic splitting}\label{EMsps}

It is interesting to investigate the mathematical structure on the Hilbert space for EM corresponding to this construction, particularly as a preparation for studying gravity.  Specifically, the construction we have given provides a family of embeddings, labeled by the total charge $Q$ in $U$, of a product of Hilbert spaces corresponding to $U$ and its exterior into the full Hilbert space:
\beq\label{EMsplit}
\bigoplus_Q \calh_{U,Q} \otimes  \calh_{\overline U_\epsilon,Q} \hookrightarrow \calh\ ; 
\eeq
here $U_\epsilon$ is an $\epsilon$-extension of the neighborhood $U$, and $\overline U_\epsilon$ is its complement.
This can be referred to as an ``electromagnetic splitting."  

To see this, first note that for an uncharged quantum field $\phi$, and neighborhood $U$, one can construct a similar embedding 
\beq\label{esplit}
\calh_{U} \otimes  \calh_{\overline U_\epsilon} \hookrightarrow \calh
\eeq
by constructing the split vacuum $|U_\epsilon\rangle$ (see \cite{Haag} and references therein; for further discussion see \cite{DoGi3,DoGi4}).  The split vacuum has the property that for operators $A$ and $A'$ which are only supported in $U$ or $\overline U_\epsilon$, respectively, 
\beq\label{splitdef}
\langle U_\epsilon|A A'|U_\epsilon\rangle = \langle 0|A|0\rangle \langle 0|A'|0\rangle\ ;
\eeq
this means that correlations between excitations in $U$ and $\overline U_\epsilon$ are removed in the split vacuum.  We also assume that a similar construction is possible for free EM fields; this is discussed in \cite{HoDa}.

In the case of a charged operator acting in $U$, the dressing must extend to infinity as in \eqref{Fline} or \eqref{gdress}.  However, if this operator is taken to be of the form \eqref{commdress}, then measurements of operators $A'$ outside $U_\epsilon$ only depend on the total charge, as well as the choice of standard dressing.  This latter choice is considered part of the definition of the electromagnetic splitting \eqref{esplit}. Specifically, consider an operator $A$ of total charge $Q$, and its correspondingly dressed version, $\hat A$; one can likewise dress $A'$ in some fashion, and for simplicity take the latter dressing to lie outside $U_\epsilon$.  
With these choices, \eqref{splitdef} generalizes to
\beq
\langle U_\epsilon|\hat A \hat A'|U_\epsilon\rangle = \langle 0|\hat A|0\rangle \langle Q|\hat A'|Q\rangle\ ,
\eeq
where the latter matrix element depends on the choice of standard dressing.

In EM, one can avoid this construction and instead construct a true splitting as in \eqref{splitdef}, by placing ``screening charges" in the region $\overline U_\epsilon \backslash U$; see {\it e.g.} \cite{HaOo} for discussion.  However, since this is problematic in gravity, we instead focus on the preceding construction.

\section{Gravitational dressings and soft charges}

A number of aspects of  the gravitational case are similar to those of the EM case, but there are important differences as well.  An important question in gravity is how much we can learn about a  matter distribution in a region by making measurements of the gravitational field of that distribution outside that region.  Specifically, in making contact with quantum information questions, we would like to understand to what extent information is localized -- can we define a localized ``gravitational qubit"? While the full theory of quantum gravity is still unknown, we can hope to infer important information by working perturbatively.  Within the context of the perturbative theory, the reason information  isn't completely localized is because of the gravitational constraints, which enforce gauge invariance and imply a certain amount of nonlocality.\footnote{In the present discussion, this manifests itself in generic failure of observables to commute at spacelike separation (see, {\it e.g.}, \cite{GiLia,GiLib,LQGST,SGalg,DoGi1}); as discussed in \cite{Giddings:2011xs} (see sec.\ 8.3), related effects appear in modification to other criteria for locality, such as cluster decomposition.}

\subsection{Dressed operators}

Consider a scalar $\phi(x)$, coupled to gravity,
\beq
\call = \frac{2}{\kappa^2} R -\hf  \left[(\nabla\phi)^2+m^2\phi^2\right]\ .
\eeq
We will work perturbatively in $\kappa=\sqrt{32\pi G}$, and expand the metric about flat space as
\beq
g_{\mu\nu}= \eta_{\mu\nu} +\kappa h_{\mu\nu}\ .
\eeq
The gauge symmetries are the diffeomorphisms, which take the  form
\beq\label{diffdef}
\delta \phi = -\kappa \xi^\mu \partial_\mu \phi+\calo(\kappa^2)\quad ,\quad \delta h_{\mu\nu}= -\partial_\mu\xi_\nu -\partial_\nu\xi_\mu+\calo(\kappa)\ .
\eeq

The local field $\phi(x)$ is not gauge invariant, and therefore is also not an observable.  Corresponding gauge invariants can be constructed, however, by dressing $\phi(x)$; we give the leading order form of these, working in the $\kappa$ expansion.\footnote{For preceding examples, see \cite{Heem,KaLigrav}.}  
These take the form
\beq\label{gengdress}
\Phi(x^\mu) = \phi(x^\mu + V^\mu(x))
\eeq
where $V^\mu$ is a functional of the metric.  
A particularly simple dressing is the gravitational line\cite{DoGi1,QGQFA},
\beq\label{Gline}
 V_\mu^\Gamma(x)= {\kappa\over 2} \int_x^\infty dx^{\prime\nu} \left\{ h_{\mu\nu}(x') + \int_{x'}^\infty dx^{\prime\prime\lambda}\left[\partial_\mu h_{\nu\lambda}(x'') - \partial_\nu h_{\mu\lambda}(x'')\right]\right\}
\eeq
where the integrals run along an arbitrary curve $\Gamma$ connecting $x$ to infinity.  This can easily be shown to transform under diffeomorphisms \eqref{diffdef} by the key relation
\beq\label{dressxm}
\delta V^\mu = \kappa \xi^\mu(x)\ .
\eeq
Given this transformation law and \eqref{diffdef}, the resulting expression $\Phi_\Gamma(x)$ is clearly diffeomorphism invariant to leading order, 
$\calo(\kappa)$.  
Equivalently, one can show that this operator commutes with the  constraints,
\beq
G_{0\mu}(x)-8\pi G\,T_{0\mu}(x)=0
\eeq
which generate gauge transformations, at this order.
Concrete expressions for commutators can be worked out as in \cite{DoGi1} by, {\it e.g.}, employing a Feynman gauge-fixing term, 
\beq
\call_{\rm gf} =-\frac{1}{\alpha\kappa^2}\frac{1}{\abs{g}^{3/2}}\left[\partial_\mu\left(\sqrt{|g|} g^{\mu\nu}\right)\right]^2\ .
\eeq

Expression \eqref{Gline} of course does not describe the most general gauge-invariant dressing.  A much broader class of dressings is of the form (for simplicity just given at $t=0$)
\beq\label{ggdress}
V_\mu(0,\vec x)=\int d^3x' {\check h}^{ij}_{\vec x}(\vec x') \gamma _{\mu,ij}(0,\vec x')\ ,
\eeq
where 
\beq
\gamma _{\mu,ij}=\frac{\kappa}{2} \left(\partial_i h_{\mu j} +\partial_j h_{\mu i}- \partial_\mu h_{ij}\right)
\eeq
is the linearized Christoffel symbol, and $\check h_{ij}$ is taken to satisfy
\beq
\partial_i\partial_j{\check h}^{ij}_{\vec x}(\vec x')= -\delta^3(\vec x'-\vec x)\ .
\eeq
When applied to the vacuum, \eqref{gengdress}, with \eqref{ggdress},  creates a $\phi$ particle together with a gravitational field determined by ${\check h}^{ij}_{\vec x}$.

There is an infinite number of such gravitational dressings at linear order in $\kappa$.\footnote{Indeed, \eqref{ggdress} may be further generalized.}
The states they create differ from one another in the radiative (source-free) part of the gravitational field.  For example, the operator \eqref{Gline} creates a gravitational field concentrated in a thin filament.  This is not the most energetically favored field state, and so will emit radiation to null infinity (${\cal I}^+$) and, for a static source in the absence of other influences, is expected to settle down to a Coulombic (linearized Schwarzschild) field\cite{DoGi1}, as in the EM case.

Also as with the EM case, special configurations with antipodal identification of the field at infinity have been considered\cite{StBMS,astrorev}, but no such identification is assumed here.  The dressing   \eqref{Gline}, where $\Gamma$ is the positive $z$ axis, illustrates this.  Again, regulation is needed in order to avoid an infinite energy configuration.  As with the EM case, we can consider regulation  with non-trivial dressing in a narrow cone extending to infinity; one way to construct such a regulated dressing\cite{GiKi} is to average the line dressing \eqref{Gline} over a small solid angle.  As with the EM case, this is expected to create (after smoothing near $r=0$) a gravitational field consisting of the Coulomb field plus a radiation field with finitely more energy, and this initial data is expected to have consistent evolution.  This construction illustrates the possible existence of a general class of field configurations not antipodally identified at infinity.\footnote{These more general configurations also modify the global version of the ``conservation laws" of \cite{StBMS}.}

\subsection{Soft charges}

The soft charges provide an interesting set of characteristics of a gravitational field.  These can be defined either at null infinity, $\cal I^{\pm}$, or at spatial infinity, $i^0$.  These are often given in a specific gauge; for example, in the gauge used in \cite{BBM}, the supertranslation and superrotation charges at $\cal I^+$ take the form (see, {\it e.g.}, \cite{astrorev})
\beq
Q_\epsilon = \frac{1}{\kappa} \int d\Omega \epsilon(\theta^A)\, \lim_{r\rightarrow\infty} \left(r h_{uu}\right)
\eeq
and 
\beq
Q_{\epsilon^A} = \frac{1}{\kappa} \int d\Omega \epsilon^A(\theta^B)\, \lim_{r\rightarrow\infty} \left( r h_{rA} +\cdots  \right)\ ;
\eeq
in the latter, $\epsilon^A$ is a vector field on the sphere, and derivative terms have been suppressed.  
These charges are, as expected, linear in the metric perturbation.  

Since dressed operators create nontrivial asymptotic fields, they should have nonzero soft charges.  For a general soft charge $Q_\iota$ at $i^0$, we find
\beq\label{softgc}
[Q_\iota,\Phi(x)] = [Q_\iota, V^\mu(x)]\,\partial_\mu \Phi(x)\ .
\eeq
The first, commutator, term on the right hand side is a c-number that corresponds to the soft charge of the field created by the dressing $V^\mu$.  This will of course depend on the specific choice of dressing $V^\mu$.

It is useful to generalize this to find the action of the soft charges on an arbitrary dressed operator.  If $A$ is an operator in the $\kappa=0$ field theory, and $V^\mu(x)$ is a dressing satisfying \eqref{dressxm}, then \cite{DoGi4} showed that the dressed operator
\beq\label{genopd}
{\hat A} = e^{i\int d^3x\, V^\mu(x)\, T_{0\mu}(x)}\ A\ e^{-i\int d^3x\, V^\mu(x)\, T_{0\mu}(x)} \  + {\cal O}(\kappa^2)\ ,
\eeq
with $T_{0\mu}$ the stress tensor,  commutes with the constraints, to leading order in $\kappa$, and so is diffeomorphism invariant to this order.
Then, \eqref{softgc} generalizes to
\beq\label{qcomm}
[Q_\iota,{\hat A}] =i\int d^3x\, [Q_\iota,V^\mu(x)]\,[T_{0\mu}(x),A]\ .
\eeq
Likewise, states can be dressed as\cite{DoGi4}
\beq\label{gstated}
|\psi\rangle \rightarrow |{ \Psi}\rangle=e^{i\int d^3x\, V^\mu(x)\, T_{0\mu}(x)}|{\psi}\rangle \ +{\cal O}(\kappa^2)\ ,
\eeq
and will have non-trivial soft charges.  Plausibly states can be constructed with all possible values of soft gravitational charges, but that will not be carefully checked here.

Since \eqref{genopd} gives the dressing of an arbitrary operator, this includes multiparticle operators.  It is also instructive to examine matrix elements of products of soft charges.  By commuting the dressing through the charges, we find expressions analogous to \eqref{qcomm},
\beq\label{prodch}
\langle\Psi' | \prod_k Q_k| \Psi\rangle = \langle \psi' | \prod_k \left\{Q_k-i\int d^3x   [Q_k,V^\mu(x)] T_{0\mu}(x) \right\} |\psi\rangle\ .
\eeq
If we can take $Q_k|\psi\rangle=Q_k|\psi'\rangle=0$, this reduces to 
\beq
\langle \psi' | \prod_k \left\{-i\int d^3x   [Q_k,V^\mu(x)] T_{0\mu}(x) \right\} |\psi\rangle\ ,
\eeq
but this simplification is potentially problematic in light of the possible existence of an infinite number of degenerate vacua acted on nontrivially by the soft charges\cite{BBM,Sachs}\cite{astrorev}.

\subsection{Decoupling soft charges}

We can now ask what information can be determined about a matter distribution, say that is localized in a neighborhood $U$, by measurements of the gravitational soft charges.  We can consider in particular states created by a dressed operator \eqref{genopd}, which could, for example, be a product of ``single-particle" operators \eqref{gengdress}.  The discussion is similar to the EM case.
 If we pick a specific dressing $V^\mu$ that is used to dress all the particles, then field observations outside $U$ can measure aspects of the charge distribution inside $U$; for an example consider dressing all the particles by gravitational lines, \eqref{Gline}.
 However, this particular choice of dressing of the operators assumes a particular correlation between the state of the gravitational field and the configuration of particles.  There are equally good dressings that do not have this correlation.  
 
A dressing that illustrates this is constructed as follows\cite{DoGi4}.  Choose a fiducial point $y\in U$, and choose some dressing $V_S^\mu(y)$ which we will call a ``standard dressing;" this could be a gravitational line \eqref{Gline}, a Coulomb dressing, or some other dressing, which satisfies the key relation \eqref{dressxm} at point $y$.  
The dressing $V^\mu(x)$ is then constructed by combining this with a gravitational line running from $x$ to $y$.  Specifically, define a generalization of the line dressing \eqref{Gline}, 
\beq
V_\mu^L(x,y)= -{\kappa\over 2} \int_y^x dx^{\prime\nu} \left\{ h_{\mu\nu}(x') - \int_{y}^{x'} dx^{\prime\prime\lambda}\left[\partial_\mu h_{\nu\lambda}(x'') - \partial_\nu h_{\mu\lambda}(x'')\right]\right\}\ .
\eeq
Then it is easily checked that 
\beq\label{compdress}
V^\mu(x)=V_L^\mu(x,y)+V_S^\mu(y) + {1\over 2}(x-y)_\nu[\partial^\nu V_S^\mu(y)-\partial^\mu V_S^\nu(y)]
\eeq
satisfies \eqref{dressxm} for arbitrary point $x$.  This dressing is analogous to the EM dressing \eqref{commdress}.  The line dressing $V_L^\mu(x,y)$ creates only excitations within $U$, if the curve between $x$ and $y$ lies inside $U$, and only the terms involving $V_S^\mu$ create excitations outside $U$.  
As before, the dressing field created by \eqref{compdress} differs from that of a collection of gravitational lines, or from {\it e.g.} a collection of Coulomb fields, by a purely radiative (sourceless) configuration of the gravitational field.

Suppose that the undressed operator $A$ is localized to $U$ ({\it i.e.} lies in the subalgebra of operators with support only in $U$).  
Then, in the commutator \eqref{qcomm}, or in a more general such commutator with $Q_\iota$ replaced by an operator outside $U$, use of the dressing \eqref{compdress} leads to a simplified expression only depending on the Poincar\'e charges of $A$.  Specifically, define the (c-number) soft charges of the standard dressing as
\beq
[Q_\iota,V^\mu_S(y)] = q^\mu_{\iota,S}\ .
\eeq
Then the commutator \eqref{qcomm} becomes
\beq
[Q_\iota,{\hat A}] = -i q^\mu_{\iota,S}(y)\, [P_\mu,A] -\frac{i}{2} \partial^\mu q^{\nu}_{\iota,S}(y)\, [M_{\mu\nu},A]\ .
\eeq
This commutator depends just on the choice of standard dressing and on the Poincar\'e charges of $A$, analogous to the dependence of  asymptotic observations of \eqref{commdress} only  on total electric charge.  Similarly, the matrix element of a product of charges, \eqref{prodch} becomes
\beq\label{prodchS}
\langle\Psi' | \prod_k Q_k| \Psi\rangle = \langle \psi' | \prod_k \left\{Q_k+i   q^\mu_{\iota,S}(y)\, P_\mu + \frac{i}{2}\partial^\mu q^{\nu}_{\iota,S}(y)\, M_{\mu\nu} \right\} |\psi\rangle\ .
\eeq
This expression tells us that measurement of a general product of soft charges is determined by the choice of standard dressing and the matrix elements of the Poincar\'e generators, along with possible soft charge correlators arising from vacuum degeneracy.

These statements lead to the following conclusions:

\begin{enumerate}

\item  Measurements at $i^0$ (and those of soft charges, in particular) can't necessarily detect details of the matter distribution (state) inside $U$.  

\item  Such measurements {\it do} detect properties of the radiation field that has been superposed on this matter distribution.

\item  We can always {\it choose} to consider initial states where the matter distribution and the radiation field are correlated in a certain way, but we can also choose states where they are correlated in a different way.  The only necessary correlation is determined by the total Poincar\'e charges $P_\mu$ and $M_{\mu\nu}$;  the asymptotic gravitational field must carry these charges.

\item  In states with specific correlations between the matter distribution and the asymptotic field, one can of course asymptotically measure those aspects of the matter distribution correlated with the asymptotic field.

\end{enumerate}

The expression \eqref{prodchS} suggests that we may asymptotically measure arbitrary correlation functions of the Poincar\'e generators $P_\mu$ and $M_{\mu\nu}$.  To be clear, these refer to the total, or center of mass (CM), momentum and angular momentum.  Of course, these generators do not all commute, but we can specify wavefunctions or states as functions of a maximally commuting set.  In addition to the Casimir $P^2$, the Pauli-Lubanski vector
\beq
W^\mu= \frac{1}{2}\epsilon^{\mu\nu\lambda\sigma}P_\nu M_{\lambda\sigma} 
\eeq
gives a second Casimir $W^2$.   We can also simultaneously diagonalize the spatial momenta, $P_i$, and one component of the Pauli-Lubanski vector, $W_z$.  In addition, there can be a large number of other quantum numbers of a general multiparticle state, which we collectively denote by $\alpha$.  Generic multiparticle  states give massive representations, $P^2=-M_\alpha^2<0$, so $W^2=M_\alpha^2 s_\alpha (s_\alpha+1)$, where $s_\alpha$ can take on values $0, \frac{1}{2}, 1, \cdots$, and the eigenvalue of a component $W_z/m$ ranges over $-s_\alpha, -s_\alpha+1,\cdots,s_\alpha$.  Thus a basis of states can be written
\beq
|\alpha, M_\alpha, s_\alpha; P_i,s_z\rangle
\eeq
where $M_\alpha$ and $s_\alpha$ are determined by $\alpha$. A generic multiparticle state associated to $U$ can be written\footnote{Note, however, that such states localized to $U$ cannot be eigenstates of $P_i$.}
\beq
|\psi\rangle = \sum_\alpha \int d^3P \sum_{s_z} \psi_\alpha(P_i, s_z) |\alpha, M_\alpha, s_\alpha; P_i,s_z\rangle
\eeq
where the sum over $\alpha$ generically also contains integrals, and is of high dimension.  

We expect measurement of arbitrary correlation functions of Poincar\'e generators to determine the dependence of the wavefunction on $P_\mu$, $W_z$.
 This of course means that single particle states can be determined at infinity\cite{DoGiunp,WDsin}.  But general states of particles associated to $U$ will be labeled by the enormous number of other quantum numbers $\alpha$.  Any states that have degenerate $M_\alpha, s_\alpha$ cannot be distinguished  by these measurements outside $U$, if they are in identical wavefunctions for the CM variables.  This is true both for measurements of the soft charges and 
for more general measurements of the gravitational field.  We leave for future exploration the study of such degeneracies, as well as the constraints on distinguishability of nearly degenerate states.  But, in this sense, asymptotic measurements are in general independent of any details of the matter distribution.    Practically, it means that information can be localized ({\it i.e.} is not measurable outside $U$) for  any degenerate states coupled to gravity, or put colloquially, there can be ``gravitational qubits."

\subsection{Information in black holes}

So far the discussion has focused on construction and observation of states on a Minkowski spacetime background.  An important question is whether states inside a black hole ``express"\footnote{This terminology is used in loose analogy with gene expression; there, the information of a gene is used in the synthesis of a gene product.  In the black hole case, information about the black hole state must play a role in determining the detailed outgoing state.} their information in a way such that they violate the basic arguments about loss of information in Hawking radiation; to resolve the associated ``unitarity crisis," this expression would have to be sufficient to transfer all information associated to the black hole to the outgoing radiation, to render the decay unitary.  This possibility has in particular been a theme in discussions of soft hair\cite{Hawk,HPS1,HPS2,astrorev,astrorevisit,HPPS}.

Construction of dressing of operators and states that describe perturbations about a black hole background is more complicated than in a flat background,\footnote{Some discussion will be provided in forthcoming work\cite{GiWe}.} and so a rigorous argument cannot yet be provided.  However, there is good reason to expect that the general features of the preceding arguments will extend to the black hole case.  Specifically, if a localized region in flat spacetime can contain information that is not observable outside the region, in particular via measurements of the gravitational field, it seems quite plausible that this statement holds also for a localized region that happens to be inside a black hole.  Indeed, if we dress states or operators inside a black hole, the corresponding dressing should extend to infinity as in the flat case.  This dressing creates a gravitational field associated with the corresponding particles.  But, that field can be augmented by an arbitrary field describing gravitational radiation.  Thus, there is little {\it required} correlation between the gravitational field seen outside the black hole, and the states inside the black hole, as in the flat case where the only required correlation is via the total Poincar\'e charges.  Elaboration of these arguments is thus expected to show that, in the same perturbative sense, information can be localized in a black hole so that it is not accessible outside.  

In particular, the preceding discussion suggests that there can be different black hole states with the same exterior gravitational field state, {\it e.g.} of the form
\beq
|K,\psi\rangle
\eeq
where $K$ labels  black hole states, and $\psi$ is the external state.  It may {\it also} be possible to choose states where the exterior is correlated with the black hole state, {\it e.g.}
\beq
|K,\psi_{K}\rangle\ .
\eeq
However, the existence of the former indicates that exterior quantum degrees of freedom, {\it e.g.} of Hawking radiation, can be entangled with the black hole states, with no outside manifestation, showing that there can be missing information.  
If this is the case, the problem of nonunitarity is expected to appear, and some other effect beyond ``soft hair" is needed to transfer information out of the black hole to unitarize evolution.  For a recent discussion of parameterization of such effects, see \cite{NVU}.

\section{Gravitational splitting}

Viewed from a ``quantum-first" perspective\cite{UQM}\cite{QFG,QGQFA}, beginning with the postulates of quantum mechanics, a key question for quantum gravity is what mathematical structure on Hilbert space is necessary to describe the foundations of the theory.  Since any such structure should presumably have a perturbative version, at least for certain states of the theory, this suggests that important clues should be found in the perturbative structure of the theory.

In most quantum theories, the underlying description involves definition of ``quantum subsystems."  This is important to provide a notion of Einstein separability, in order to describe physics.\footnote{For some further discussion, see \cite{QFG,QGQFA}.}  For finite quantum systems, subsystems are determined by tensor factorization of the Hilbert space.  For quantum field theories, quantum subsystems can be thought of as determined by local subalgebras of the algebra of observables, associated with spacelike-separated regions.  Alternatively, via the construction of  the split vacuum described in sec.~\ref{EMsps}, subsystems in field theory can be defined in terms of embeddings of product Hilbert spaces into the full Hilbert space, of the form \eqref{esplit}.

As is discussed in \cite{SGalg,QFG}, neither of these approaches appears to define quantum subsystems for quantum gravity.  However, the case of electromagnetic splitting, described above, suggests a way to proceed for gravity.  Specifically, the dressing construction \eqref{genopd}, \eqref{gstated}, \eqref{compdress} makes observations of field configurations outside a neighborhood $U$ depend only on the total Poincar\'e charges of the excitations inside the neighborhood.  This suggests that it defines a ``gravitational splitting," of the form (compare \eqref{EMsplit})
\beq\label{Grspl}
\bigoplus_{P_\mu, s, s_z}  \calh_{U,P_\mu,s, s_z} \otimes  \calh_{\overline U_\epsilon,P_\mu,s, s_z} \hookrightarrow \calh
\eeq
with Hilbert space factors labeled by the simultaneously-measurable Poincar\'e charges.   A subtlety, mentioned above, is  that states associated to $U$ (precisely:  gotten by acting with operators restricted to $U$, on the split vacuum $|U_\epsilon\rangle$) are not allowed to be eigenstates of the momenta.  So, while \eqref{Grspl} represents certain aspects of the structure, more refinement may well be needed to give a careful mathematical definition of the subsystem structures introduced by the standard dressing procedure \eqref{genopd}, \eqref{gstated}, \eqref{compdress}.  We would also like to extend this construction to higher orders in perturbation theory, as well as to study its nonperturbative extension; the latter, in particular, is expected to have important subtleties, particularly associated with strong gravitational fields \cite{QFG,QGQFA}.

\section{Acknowledgements}

I thank G. Compere, M. Henneaux, M. Perry, A. Strominger, S.~Weinberg, and A. Zhiboedov for valuable conversations, and the CERN theory group, where this work was carried out, for its hospitality.  This material is based upon work supported in part by the U.S. Department of Energy, Office of Science, under Award Number {DE-SC}0011702.

\bibliographystyle{utphys}
\bibliography{grsp}

\end{document}